# Comparative analysis of computational approaches for predicting Transthyretin (TTR) transcription activators and human dopamine D1 receptor antagonists

Mariya L. Ivanova[1,*], ID, Nicola Russo[1], ID , Gueorgui Mihaylov[2], ID and Konstantin Nikolic[1], ID

Author affiliations

[1]School of Computing and Engineering, University of West London, London, UK

[2]Haleon, London, UK

*Corresponding author mariya.ivanova@uwl.ac.uk

## Abstract

The study expands the application of scikit-learn-based machine learning (ML) to the prediction of small biomolecule functionalities based on Carbon 13 isotope (13C) NMR spectroscopy data derived from Simplified Molecular Input Line Entry System (SMILES) notations. The methodology previously demonstrated by predicting dopamine D1 receptor antagonists was upgraded with the addition of new molecular features derived from the PubChem database. The enhanced ML model obtained 75.8% Accuracy, 84.2% Precision, 63.6% Recall, 72.5% F1-score and 75.8 % ROC, when is trained on 25,532 samples and tested on 5,466 samples. To evaluate the applicability of the methodology for a variety of case studies, a comparison was conducted between the prediction capabilities of the ML models based on the human dopamine D1 receptor antagonists and on the neuronal Transthyretin (TTR) transcription activators. Since the TTR bioassay did not contain the required number of samples for comparison, the results were obtained hypothetically. Gradient Boosting classifier was the optimal model for TTR transcription activators, achieving hypothetical 67.4% Accuracy, 74.0% Precision, 53.5% Recall, 62.1% F1-score, 67.4 % ROC, if it could be trained with 25,532 samples and tested with 5,466 samples. In addition to the main study, to the attention of those interested in neuronal TTR, the CID_SID ML model has been developed to predict whether a compound, initially designed for another purpose, possesses TTR transcription activation capabilities. This ML model was based solely on its PubChem CID and SID and achieved 81.5% Accuracy, 94.6% Precision, 66.8% Recall, 78.3% F1-score, 81.5 % ROC.

## Introduction

Nuclear Magnetic Resonance (NMR) Spectroscopy has been considered as one of the most powerful and informative techniques currently available for small biomolecule analysis (1). It provides detailed information about the arrangement of atoms within a molecule; identification of functional groups; three-dimensional arrangement of atoms in space; dynamics and interaction with other molecules; quantitative analyses (2). Enabling computers to learn from data through analysis and eliminating explicit programming, ML has been applied to data obtained via NMR spectroscopy such as employing message passing neural networks (MPNNs) to predict 13C NMR chemical shifts in small molecules (3); decreasing the ML errors of the NMR spectroscopy chemical shift predictions by using Deuterated chloroform (4); Computational structural elucidation aided by NMR spectroscopy and machine learning techniques (5); combining AlphaFold and chemical shift prediction in a deep learning model for faster protein NMR assignment (6); built from over 8,300 carbon atoms in diverse environments and without limitations on molecular complexity, a general-case neural network model was created for predicting (estimating) 13C NMR spectra (7).

Beyond ML, a range of computational methods listed below have been employed to predict NMR (8, 9), such as:

(i) Computational "ab initio" techniques, meaning they operate "from first principles," determine a system's properties using only fundamental laws, in contrast to methods relying on empirical data. One such application is the development of a 13C NMR chemical shift prediction procedure for small molecule structure elucidation, which employed gauge invariant atomic orbitals with density functional theory and incorporated empirical systematic error correction terms (10).

(ii) Additive increment-based approaches represent the initial methods for predicting chemical shifts where the core atomic shift has been subsequently modified by positive or negative contributions from attached substituents. Such rules, for instance, have been implemented in the CASPER program (11) to predict 13C, 1H, and 15N shifts of polysaccharides.

(iii) Hierarchical Organisation of Spherical Environments (HOSE) code characterises an atom's molecular environment by iteratively describing concentric spheres of neighbouring atoms. Essentially, this approach operates as a nearest neighbour search, where the HOSE code serves as the metric for defining structural similarity (9)

The online tool NMRDB (12) employed in the study to convert SMILES into 13C NMR spectroscopy data has been based on HOSE code and databases of existing NMR spectra and chemical structures reviewed by a board of reviewers (13). Recognising that NMRDB-derived information represents a first approximation compared to experimental data, its employment was sufficient for the demonstration of the proposed methodology idea within the current and previous studies.

The presented research had to solve two tasks: first, to assess the applicability of a pre-existing methodology (14) to a different case study; and second, to determine how the inclusion of molecular features would affect the accuracy of the investigated machine learning model. The existing methodology was an ML approach based on NMR spectroscopy data derived from SMILES notations, demonstrated by the case study on human dopamine D1 receptor antagonists, PubChem AID 504652 (15), which data was provided by PubChem, the world`s largest database for freely available chemical information (16). The core idea of the existing ML approach was to leverage the relationship between a small biomolecule's carbon skeleton and the information provided by 13C NMR spectroscopy. The NMR spectrum, with its peaks, indicates atomic bonds and ligands and elucidates the molecular environment of the carbon atoms. This data was numerically encoded for machine learning, enabling its association with labels. So, it was anticipated and respectively demonstrated that 13C NMR spectroscopy could effectively contribute to the ML-based prediction of small biomolecule functionality.

The new case study, PubChem AID 1117267 bioassay (17) used for the comparison was also provided by PubChem. It was focused on TTR transcription activators containing 91,943 samples, 1,155 of which were defined as TTR activators. Data was obtained by High-throughput screening (HTS), which is an automated method in drug discovery and toxicology that rapidly tests vast compound libraries for biological activity using robotics and automated systems for parallel processing and analysis. The bioassay was conducted to identify small biomolecules activating neuronal TTR transcription. Compounds are tested in concentration of 16.7 uM. For more information about the screening protocol, please refer to the bioassay documentation [17].

The protein TTR was called “the servant of many masters”. The primary functions of TTR are the transport of thyroid hormones and retinol through the bloodstream and cerebrospinal fluid (18). However, its stability is critical, and misfolding can lead to amyloid disease, causing amyloidosis of nerves, ligaments, heart and arterioles (19) that could lead to polyneuropathy where multiple peripheral nerves become damaged (20), and/or cardiomyopathy affecting the heart muscles (21). Additionally, TTR has been reported to play an independent neuroprotective role in both the peripheral and central nervous systems, contributing to nerve function and repair (22, 23, 24). Using TTR Knockout mouse (25), it has been found out that the absence of transthyretin leads to increased levels of neuropeptide Y (NPY) in the brain and peripheral nervous system. Despite the need for additional research to fully elucidate the specific effects in TTR knockout mice, the study confirmed the presence of alterations in feeding behaviour, anxiety levels, stress responses, metabolic parameters, and potential changes in nerve function and repair (26).

So, overall, the protein TTR is a subject of research for many scientists and clinicians due to its crucial roles in the body and its association with a significant and debilitating group of diseases known as transthyretin amyloidosis (ATTR).

Regarding the second objective of the presented research, which expanded the study beyond the comparison, molecular features obtained from PubChem were added to the NMR spectroscopy data to explore their influence on the ML models. A similar approach, which has improved the ML model, has been performed in studies regarding G9a (27, 28). These molecular features were as follows:

(i) Rotatable Bond Count (RBC) is determined by counting the single bonds that are not part of a ring and connect two non-hydrogen atoms that are also not at the end of a chain.
(ii) Hydrogen Bond Acceptor Count (HDAC) refers to the number of hydrogen bond acceptors present in the given small biomolecule.
(iii) Hydrogen Bond Donor Count (HBDC) refers to the number of hydrogen bond donors present in the given small biomolecule.
(iv) Topological Polar Surface Area is computed as the surface sum over polar atoms in the molecule (29)
(v) XLogP3 (XL) represents a predicted octanol-water partition coefficient (30).
(vi) Molecular Weight (MW) is defined as the sum of the mass of all constituent atoms (31).

Furthermore, an additional ML application was developed for researchers interested in TTR transcription activators aimed to predict whether a small biomolecule, initially designed for another purpose, could also function as a TTR transcription activator. This ML prediction was based exclusively on PubChem CID and SID data of the small biomolecule, using the methodology (32) that has previously demonstrated success in predicting the case studies related to CHOP inhibitors, dopamine D1 receptor antagonist, dopamine D3 receptor antagonist, TDP1 inhibitors, M1 muscarinic receptor antagonists, Rab 9 promoter activators and G9a enzyme inhibitors While identifiers are generally not used in ML, PubChem identifiers can be an exception. This is because their generation process employs an algorithm that considers the structure and similarity between substances and compounds (33). For more details, please refer to the parent study (32).

Last, but not least, since the ML model of interest was classification, i.e. heavily dependent on the balance between the classes, the severe imbalance between the active and inactive biomolecules in the TTR bioassay dataset was handled through three steps. One of these steps incorporated PubChem AID 1996 bioassay focused on water solubility of small biomolecules (34), whose dataset was used as a sieve for the reduction of the small biomolecules without considering the level of aqua solubility of the samples.

## Methodology

The methodology followed whose applicability exploration was the aim of the research. For this purpose, the CID, SMILES notations, and labels were extracted from the PubChem AID 1117267 bioassay dataset, which contains 91,909 rows of small biomolecules and nine columns of features describing these molecules. The samples detected as TTR transcription activators were 1,155, called Active and 90,755 detected as Inactive compounds. The severe imbalance between the active and inactive compounds was initially decreased by merging the dataset of bioassay AID 1117267 with the dataset of the PubChem AID 1996 bioassay. Since the latter contained 57,856 samples of small biomolecules, this margin of both datasets played the role of a sift, reducing the number of inactive compounds. Further, after shuffling of the inactive compounds only every eighth sample was retained. The reduced inactive compounds, 2,023 samples, were then concatenated with all active compounds, i.e. 1,155 samples from the PubChem AID 1117267 bioassay dataset and the final dataset with 3,177 samples in total was obtained. Further, the SMILES notations of the resulting dataset were converted into numerical spectroscopy data employing the online tool NMRDB that has been designed for this purpose. The process requested each SMILES to be uploaded on the website where the NMRDB tool will generate 13C NMR spectroscopy prediction, which was used to obtain the necessary data for ML. The scale of the chemical shifts, i.e. the pics, started from 0 and reached beyond 200. The scale was deviated by natural numbers, which defined subranges. These subranges were used as a feature and generated the columns in the data frame. The presence of a pick or picks in each range was counted and placed as a value for the corresponded subrange of the relevant small biomolecule. Once all SMILES notations were converted to 13C NMR spectroscopy data, the dataset was merged with the labels based on the SMILES notations. In this way, since the canonical SMILES are unique, misconnection between isomers was avoided.

To ensure the reliability of the evaluation, an equal number of samples from each class was extracted and the testing set created. The remaining samples were balanced with oversampling. Although SMOTE is the technique that is generally recommended for balancing of data because it is expected to reduce overfitting and improve generalisation, simulations for this study showed that Random Over Sampler was more effective. This involved increasing the minority class size to match the majority class by duplicating minority class samples. The balanced dataset was used for ML training with the RFC, DTC, GBC, and SVC estimators provided by the scikit learn ML library. The results of prediction were evaluated by the classification metrics, which are based on the four possible scenarios: true positive (TP) and true negative (TN), where the prediction corresponds to the actual values, and false positive

(FP) and false negative (FN), where they do not. So, the evaluation was done based on the metrics:

(i) Accuracy – the total of correct predictions to the total of all predictions.
(ii) Precision – true positive out of the total of all positive predictions.
(iii) Recall – true positive out of all positive instances, giving the sensitivity of the ML model
(iv) F-1 score – indicates the balance between precision and recall
(v) ROC (Receiver Operating Characteristic) shows the true positive against the false positive when the threshold varies.

The comparison of the results ranked the most optimal estimator for the case. The chosen ML model was scrutinised for overfitting, tracing for the deviation between training and testing accuracy to be smaller than 5%.

To ensure fairness in comparison between two cases, data set used for the ML model based on human dopamine D1 receptor antagonist data was reduced to the same number of active and inactive small biomolecules used in the TTR transcription activators prediction. Then, the difference in accuracy between the ML model based on the reduced D1 receptor antagonist dataset and the dataset used in the previous study was used to calculate how much the accuracy had grown due to the increase in dataset samples. It was hypothesised that the existence of the same number of samples for the TTR case would increase the accuracy of the ML model by the same percentage. ML with the enriched dataset was performed in the same manner explained above for the dopamine D1 receptor antagonist and TTR transcription activators.

Principal component analysis (PCA) was applied with the intention to reduce the noise of data by identifying the highest variance in the data, and due to effective filtering of the noise, data will become clean, and the ML model will be able to learn the actual underlying patterns more effectively.. Also, the reduction of dimensionality is expected to handle the data sparsity, helping the ML model to find a meaningful relationship. Last but not least, PCA removes multicollinearity, which can lead to more reliable metrics.

During the development of the prior study (14), which demonstrated the prediction of human dopamine D1 receptor antagonists, it was observed that the model depended on the number of samples. To ensure a fair comparison in the current study, ML based on dopamine D1 receptor antagonists was conducted, using the same number of samples as the ML prediction of TTR transcription activators. Moreover, in the current study, the molecular features were added to both the reduced dataset and the entire dataset for dopamine D1 receptor antagonists. he former was used for the purpose of comparison, and the latter was used to explore the effect that molecular features have on accuracy in general. The methodology is illustrated in Figure1.

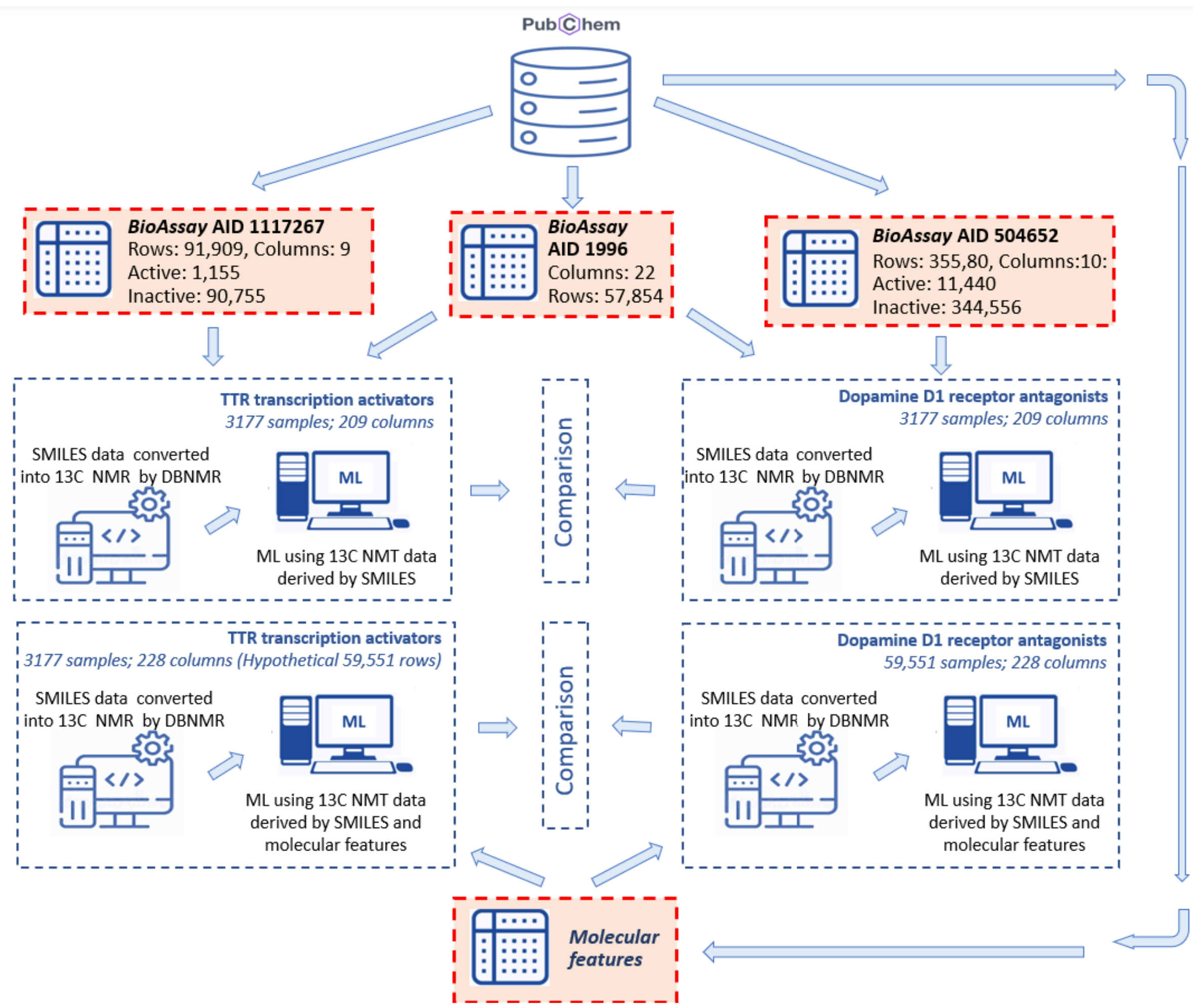

Figure 1. Methodology of comparison of dopamine D1 receptor antagonist and TTR activators based on 13C NMR spectroscopy data and molecular features.

## Results and discussion

The results of comparison of the ML models are summarised in Table 1, providing connection with the relevant tables or figures in the electronic supplementary material (ESM).

Leveraging the methodology outlined above, the initial collection of inactive compounds from the TTR bioassay was narrowed down to 2,023 small biomolecules. Combining this reduced set of inactive samples with the full set of 1154 active compounds resulted in a data set with a total of 3177 rows and 209 columns. From this dataset, a set of 340 samples per class was extracted for testing. The remaing samples, after balancing their label proportion with SMOTE, were allocated for training. As a result of this, the ML models were trained with 2497 samples and tested with 680 samples. From the classifiers listed above, GBC was the optimal estimator, achieving 56.8% accuracy, 59.2% precision, 43.5% recall, 50.2% F1-score, 56.8 % ROC (Table ESM1) and a five-fold cross-validation score of 0.651 with 0,02 standard deviation (Table ESM2). The deviation between the ML model performance and the cross validation implied overfitting, which was confirmed by comparing the training and testing accuracies (Figure ESM1). Applying PCA slightly improved the accuracy to 58.4%, precision to 65.4% and ROC to 58.4.%. However, the low recall of 35.3% decreased the F1-score to 46.1% F1-score, which was an indicator of an increase of the bias of the ML model towards the majority class (the inactive compounds), and thus, the ML model can miss crucial instances of the minority class (the active compounds) (Table ESM3). The increased restfive-fold cross-validation score of 0.745 with 0,04 standard deviation (Table ESM4) compared to the single ML model accuracy of 58.4% (Table ESM3) suggested that the ML model was overfitted, which was subsequently confirmed by tracing the deviation between train and test accuracy (Figure ESM2).

To perform a fair comparison between the two case studies, based on the random principle, equal samples of 1,948 inactive and 1,093 active compounds were extracted from both the dopamine D1 receptor antagonist and TTR transcription activator case studies. After the rows of the dopamine D1 receptor antagonist dataset were reduced to correspond to the TTR dataset, ML was performed in the same manner as it was done for the TTR transcription activators case. The optimal estimator was SVC, which obtained 65.7 % accuracy, 70.3% precision, 54.4 % recall, 61.4% F1-score, 65.7 % ROC (Table ESM5) and a five-fold cross-validation score of 0.775 with 0.02 standard deviation (Table ESM6). Expectedly, because the difference between the single and cross-validation score was significant, the ML model was overfitted (Figure ESM3). However, since the ML model based on full set of the available data for the dopamine D1 receptor antagonist was not overfitted [see the original research (14)] it was hypothesised that the increase of the samples of the TTR case equal to the same amount of samples used in the dopamine D1 receptor case, i.e. 10,542 active and 46,496 inactive compounds, would improve the TTR ML model with the same percentage as the accuracy of the dopamine D1 ML model. So, since the metrics of the ML model predicting the dopamine D1 receptor antagonist were accuracy of 71.5 %, precision of 77.4%, recall of 60.6 %, F1-score of68.0 % and ROC71.5% (Table ESM9) and a five-fold cross-validation score of 0.748±0.003 standard deviation (Table ESM10) the percentage of improvement of the ML model based on the increase of the number of samples was calculated as follow: the accuracy of the ML model with reduced rows plus the accuracy of the ML model with reduced rows multiplied by the unknown percentage should be equal to the accuracy of the ML model with non-reduced rows (Figure 2). The increase of 5.8 in the accuracy of the ML model due to an increase in the number of samples was 8.85%. So, it was hypothesised that the increase the number of samples for TTR case to the level of samples for the D1 case, i.e. 57,038 compounds would increase the accuracy of the ML model that predicts TTR transcription activators to 61.82%.

In order to improve the performance of the ML models, their complexity was mitigated by the PCA dimensionality reduction. For the TTR transcription activators case (Table ESM1 and Table ESM2 without PCA and Table ESM3 and Table ESM4 with PCA) based on the difference between the single ML metrics and the cross-validation score can be concluded that the ML models were overfitted. This was confirmed by scrutinising them for overfitting, where the difference between training and testing accuracy was higher than 5 % since the beginning, i.e. max_depth=1. For the ML model without PCA, the training accuracy was 62.8% while the testing accuracy was 54.9% (Figure ESM1). For the ML model with PCA, the training accuracy at max_depth=1 was 69% compared to the testing accuracy of 57.9% (Figure ESM2). Similarly to the ML models predicting TTP transcription activators, the presence of PCA did not improve the ML models with a reduced dataset.

Accuracy of 65.7% of the D1 ML model based on the reduced set of data

Accuracy of 71.5% of the D1 ML model based on the full set of data

Calculate the increase of accuracy of the D1 ML model due to samples increase

65.7 + 65.7.x/100 = 71.5

x = (71.5*100 -100*65.7)/65.7 = 8.83;

8.83*65.7/100 = 5.8

65.7 + 5.8 = 71.5 (dopamine D1 case`s accuracy)

Calculating the hypothetical accuracy of the TTR ML model if the dataset were big as the entire dopamine D1 dataset. The increase of accuracy was added to the accuracy of 56.8% of TTR case study:

8.83*56.8/100 = 5.01

56.8+5.01 = 61.8* (Hypothetical accuracy)

Figure 2. Calculating the hypothetical accuracy of ML model predicting TTR activators based on 13C NMR spectroscopy data.

The accuracy without PCA was 65.7% (Table ESM5) with 0.775 cross-validation score (Table ESM6) compared to the accuracy with PCA that was 64.7% (Table ESM7) with a cross-validation score of 0.748 (Table ESM8). The addition of samples to the dataset used for ML without PCA dimensionality reduction improved the ML model performance to 71.5% of accuracy (Table ESM9) with cross validation of 0.749 (Table ESM10), and significant overfitting has not been detected (Figure ESM3). The increase in the number of samples in the PCA dimensionality reduced case, on the other hand, did not increase the accuracy. It was accuracy of 64.7% (Table ESM7) with a five-fold cross-validation score of 0.748 (Table ESM8) and become 64.2% accuracy (Table ESM11) with 0.683 five-fold cross-validation score (Table ESM12), however the decrease of the difference between the single ML model accuracy (Table ESM11 ) and the five-fold cross-validation score (Table ESM12) was an indicator that due to the application of PCA the ML model became more robust than the ML model based on reduced dataset (Figure ESM4).

The next major step in the study, the addition of above-mentioned molecular features to the 13C NMR spectroscopy data, improved the ML models` performance in both the TTR transcription activators and dopamine D1 receptor antagonists cases. Regarding the TTR transcription activators dataset, which had 3177 samples, merging it with the molecular data led to a decrease in the dataset to 3,041 samples (1,948 inactive and 1,093 active compounds). The optimal ML model for this case was GBC, achieving 67.1% accuracy, 74.0% precision, 52.6% recall, 61.5% F1-score, 67.1 % ROC (Table ESM13) and a five-fold cross-validation score of 0.693 with 0,019 standard deviation (Table ESM14). The scrutiny for overfitting revealed the optimal performance of GBC was at max_depth=2. At this depth, the model achieved 62.4%accuracy, 66.7% precision, 49.4% recall, 56.8% F1-score and 62.4% ROC (Figure ESM5). The PCA dimensionality reduction in case of TTR dataset with 13CNMR spectroscopy data and molecular features worsened the performance of the GBC ML model to 60.4% accuracy, 62.5% precision, 52.1% recall, 56.8% F1-score, 60.4 % ROC (Table ESM15) and five-fold cross-validation score of 0.717 with ±0.013 standard deviation (Table ESM16). Regarding the ML models predicting whether a compound is a dopamine D1 receptor antagonist, the ML model accuracy increased from 65.7% (Table ESM 5) to 75.1% (Table ESM17) with a five-fold cross-validation score of 0.788≠0.013 standard deviation (Table ESM18). The dimensionality reduction with PCA in this case worsened the results to an accuracy of 70% (Table ESM19) and a five-fold cross-validation score of 0.835±0.038 standard deviation (Table ESM20). The increased of the gap between the accuracy of the single ML model and the cross-validation score indicated for increase in overfitting.

As mentioned above, since there was a significant quantity of real data about the dopamine D receptor antagonist, the addition of molecular features was explored for the whole dataset as well, with and without PCA. Expectedly, the increase in the number of samples improved the performance of the optimal ML model GBC, obtaining 75.8% accuracy, 84.2% precision, 63.6% recall, 72.4% F1-score, 75.8 % ROC (Table ESM21) and five-fold cross-validation score 0.761 ≠0.0039 (Table ESM22). However, the increase in the number of samples in case of dimensionality reduction did not lead to substantial changes in the ML model metrics (Table ESM23 for a single ML model and Table ESM24 for five-fold cross-validation) (Figure ESM6).

Hyperparameter tuning for each of the cases was performed by Optuna, but the suggested hyperparameters did not improve the ML models. For more details regarding the hyperparameter tuning, please refer to the code provided on GitHub. So,ooverall, it can be observed that, in both study cases, i.e. ML for predicting whether a compound is a human dopamine D1 receptor antagonist or a TTR transcription activator, the addition of the aforementioned molecular features to the 13CNMR spectroscopy dataset improved the performance of the ML models. The total number of features was 227. The followed investigations for overfitting revealed that the optimal ML model, predicting whether a compound is a TTR transcription activator was GBC with default hyperparameter features and max_depth=3, achieving accuracy 75.8%, precision 84.2%, recall 63.6%, F1-score 72.5%, ROC 75.8%. The ML model was trained on 25.532 samples and tested on 5,466 samples. The optimal ML model, predicting whether a compound is a TTR transcription activator was GBC with default hyperparameter features and max_depth=2, achieving accuracy 67.4%, precision74%, recall 53.5%, F1-score 62.1%, ROC 67.4%. The ML model was trained on 3,216 samples and tested on 680 samples.

Table 1. Initial ML results obtained before estimating the generalisation through scrutinising for overfitting. The TTR ML values marketed with * hypothetical value calculated theoretically based on the percentage of increase of metrics for dopamine D1 receptor antagonists based on increase of the samples.

| | **Train with 27,756 samples** | | **Test with 5,466 samples** | |
|---|---|---|---|---|
| **Case study & dataset** | **Accuracy [%]** | **Cross-val.** | **Accuracy [%]** | **Cross-val.** |
| | ***GBC*** | | | |
| **TTR & no PCA** | 56.8 (Table ESM1) | 0.651 (Table ESM2) | 61.82* | - |
| | ***GBC*** | | | |
| **TTR & with PCA** | 58.4 (Table ESM3) | 0.745 (Table ESM4) | - | - |
| | ***GBC*** | | | |
| **TTR & no PCA & PubChem mol. data** | 67.1 (Table ESM13) | 0.693 (Table ESM14) | 67.1* | - |
| | ***SVC*** | | | |
| **TTR & with PCA & PubChem mol. data** | 65.0 (Table ESM 15) | 0.695 (Table ESM 16) | - | - |
| | ***SVC*** | | ***SVC*** | |
| **D1 & no PCA** | 65.7 (Table ESM5) | 0.775 (Table ESM6) | 71.5 (Table ESM9) | 0.749 (Table ESM10) |
| | ***SVC*** | | ***SVC*** | |
| **D1 & with PCA** | 64.7 (Table ESM7) | 0.7485 (Table ESM8) | 64.2 (Table ESM11) | 0.683 (Table ESM12) |
| | ***GBC*** | | ***GBC*** | |
| **D1 & no PCA & PubChem mol. data** | 75.1 (Table ESM17) | 0.788 (Table ESM18) | 75.8 (Table ESM21) | 0.761 (Table ESM22) |
| | ***GBC*** | | ***SVC*** | |
| **D1 & with PCA & PubChem mol. data** | 70.4 (Table ESM19) | 0.835 (Table ESM20) | 74.3 (Table ESM23) | 0.793 (Table ESM24) |

The additional computational approach, CID_SID ML model, developed to predict the TTR transcription activators achieved initially accuracy of 81.5%, precision of 94.6%, recall of 66.8%, F1-score of 78.3% and ROC of 81.5% with GBC (Table ESM 25) and cross-validation score 0.8458 with ≠0.0085 standard deviation (Table ESM 26). Optuna performed the hyper-parameter tuning of GBC. However, the suggested best parameters, which were results of one hundred study trials, did not improve the ML model. On the other hand, the scrutiny for overfitting revealed that the optimal max_depth=2(Figure ESM7). The final model, with default hyperparameter values and max_depth=2, achieved an accuracy of 81.5%, precision of 94.6%, recall of 66.8%, F1-score of 78.3%, and an ROC of 81.5%. The ML model was trained with 3.358 and tested with 680 samples. Regarding the CID_SID ML model predicting dopamine D1 receptor antagonists, the ML model has already been developed (14), obtaining an accuracy of 80.2%, precision of 86.3%, recall of 70.4%, F1-score of 77.6%, and ROC of 79.9% when trained with 19,438 samples and tested with 4,723 samples.

## Limitations

The dimensionality reduction with PCA was performed based on the suggested number of components based two methods: the first was based on the cumulative explained variance ratio, and the second employed Minka's MLE to estimate the number of dimensions (35). The number of the component for the different cases were, as follow:

(i) 209 features of dataset with 3177 samples, used to predict TTR transcription activators, was dimensionality reduced to 186 features

(ii) 209 features of dataset with 3177 samples, used to predict dopamine D1 receptor antagonists, was dimensionality reduced to 186 features

(iii) 228 features of dataset with 3177 samples, used to predict TTR transcription activators was dimensionality reduced to 186 features

(iv) 228 features of dataset with 3177 samples, used to predict dopamine D1 receptor antagonists was dimensionality reduced to 186 features

(v) 228 features of dataset with 49,551 samples, 59,551 sample, used to predict dopamine D1 receptor antagonists was dimensionality reduced to 187 features.

However, the number of PCA components could be reduced by other technique, which were not explored in the study. For the explored variants, only (iv) Improved the accuracy. Since the time needed for training was not significant, the reduction of the time needed for ML was not taken into consideration.

Hyperparameter tuning was performed by Optuna. On the one hand, the number of studies were 5 and 10, but their increase is expected to bring better results. On the other hand, there are other hyperparameter technique which can be explored in order to improve the performance of the ML models. The same is with the choice of classification estimators or balancing technique. In both cases there are room for more exploration in this direction, but the paper aim of the paper was to present the idea of the approach, the chosen techniques were sufficient for the presentation of the idea of the approach, however, there are endless options for exploration of the combination of the ML techniques.

The dataset used to predict TTR transcription activators contained insufficient number of samples. The number of samples was insufficient for the explored approach, however techniques, such as qHTS, which was used for the bioassays, has the potential to provide the necessary amount of data.

## Conclusion

The lower number of samples used for training and testing of the ML model to predict the human dopamine D1 receptor antagonists achieved lower results than the ML model trained and tested with the full sets of small biomolecules. This scalability dependence was used to calculate the hypothetical accuracy for predicting the TTR transcription activators. However, the big difference of the ML models` results for the small dataset of these two case studies implied that the ML approach is a case study dependent, so that the hypothetical accuracy is recommended to be tested with real data. Given that such data is not available, the investigation of the ML approach reminded open for exploration with other bioassays that have significant number of samples. On the other hand, the addition of atomic features of the biomolecule for both bioassays improved the ML models except the variant with atomic features and a full dataset on human dopamine D1 receptor antagonists. It performed worse than the ML model with a reduced number of features. The hypothesis is that the model's complexity obstructs it from achieving better results.

The computational approach developed beyond the main course, i.e. the CID_SID ML model, offered a cost-effective and rapid way to gain early knowledge about potential side effects of drug candidates with respect to TTR transcription activators.

Although the utilized 13C NMR spectroscopy data is typically classical, the encoded into it information corresponds to the interaction of the quantum mechanical nature of matter and the light, providing deeper insights of the interaction considered in the case studies.

## Scientific Contribution

- A step forward toward development of universal model for predicting functionalities of small biomolecules based on NMR spectroscopy data through addition of molecular features that increased the ML model accuracy with 8%.
- Developing a CID_SID ML model for TTR transcription activators that can predict with 80.2%accuracy if a small biomolecule would cause a TTR transcription activation, that in turn could be useful for predicting of such a side effect.

## Author Contributions

MLI, NR, GM and KN conceptualized the project and designed the methodology. MLI, NR and GM processed the data. MLI and NR wrote the code. KN supervised the project. All authors were involved with the writing of the paper.

## Data and Code Availability Statement

The raw data used in the study is available through the PubChem portal:
https://pubchem.ncbi.nlm.nih.gov/

The code generated during the research is available on GitHub:
https://github.com/articlesmli/NMR_ML_TTR_D1.git

## Conflicts of interest

There are no conflicts to declare.

## Acknowledgements

MLI thanks UWL Vice-Chancellor's Scholarship scheme for their generous support. We sincerely thank to PubChem for providing access to their database. This article is dedicated to Luben Ivanov.

# Supplementary materials

## Tables

Table ESM1. Results of the ML metrics regarding the TTR case performed without dimensionality reduction (GitHub file: https://github.com/articlesmli/NMR_ML_TTR_D1/blob/main/TTR/4.1.ML_notPCA.ipynb )

| 1.Algorithm | 2.Accuracy | 3.Precision | 4.Recall | 5.F1 | 6.ROC |
|---|---|---|---|---|---|
| GradientBoost | 0.568 | 0.592 | 0.435 | 0.502 | 0.568 |
| SVM | 0.560 | 0.596 | 0.374 | 0.459 | 0.560 |
| K-nearest | 0.550 | 0.596 | 0.312 | 0.409 | 0.550 |
| RandomForest | 0.538 | 0.691 | 0.138 | 0.230 | 0.538 |
| Decision | 0.521 | 0.527 | 0.403 | 0.457 | 0.521 |

Table ESM2. A five-fold cross-validation for the TTR case performed without dimensionality reduction (GitHub file: https://github.com/articlesmli/NMR_ML_TTR_D1/blob/main/TTR/4.1.ML_notPCA.ipynb )

| 1.Algorithm | 2.Mean CV Score | 3.Standard Deviation | 4.List of CV Scores |
|---|---|---|---|
| RandomForest | 0.8637 | 0.0612 | [0.7923, 0.8039, 0.8559, 0.9465, 0.9198] |
| Decision | 0.7573 | 0.0352 | [0.7285, 0.7221, 0.737, 0.8083, 0.7905] |
| SVM | 0.7288 | 0.0307 | [0.6899, 0.7043, 0.7311, 0.7786, 0.74] |
| K-nearest | 0.6875 | 0.0431 | [0.6543, 0.6256, 0.6895, 0.7355, 0.7325] |
| GradientBoost | 0.6509 | 0.0195 | [0.6365, 0.6627, 0.6196, 0.6686, 0.6672] |

Table ESM3. Results of the ML metrics regarding the TTR case performed with dimensionality reduction with PCA (GitHub file: https://github.com/articlesmli/NMR_ML_TTR_D1/blob/main/TTR/4.2..ML_withPCA.ipynb )

| 1.Algorithm | 2.Accuracy | 3.Precision | 4.Recall | 5.F1 | 6.ROC |
|---|---|---|---|---|---|
| GradientBoost | 0.584 | 0.654 | 0.356 | 0.461 | 0.584 |
| SVM | 0.572 | 0.626 | 0.359 | 0.456 | 0.572 |
| Decision | 0.563 | 0.597 | 0.388 | 0.471 | 0.563 |
| K-nearest | 0.534 | 0.551 | 0.368 | 0.441 | 0.534 |
| RandomForest | 0.531 | 0.769 | 0.088 | 0.158 | 0.531 |

Table ESM4. A five-fold cross-validation for the TTR case performed with dimensionality reduction achieved with PCA (GitHub file: https://github.com/articlesmli/NMR_ML_TTR_D1/blob/main/TTR/4.2..ML_withPCA.ipynb )

| 1.Algorithm | 2.Mean CV Score | 3.Standard Deviation | 4.List of CV Scores |
|---|---|---|---|
| RandomForest | 0.8833 | 0.0568 | [0.8145, 0.8262, 0.8826, 0.9554, 0.9376] |
| Decision | 0.7834 | 0.0361 | [0.7329, 0.7533, 0.786, 0.8187, 0.8262] |
| GradientBoost | 0.7448 | 0.0337 | [0.6988, 0.7147, 0.7489, 0.786, 0.7756] |
| SVM | 0.7374 | 0.0391 | [0.7122, 0.6895, 0.7207, 0.7964, 0.7682] |
| K-nearest | 0.6661 | 0.0458 | [0.6128, 0.6196, 0.6627, 0.7266, 0.7088] |

Table ESM5. Results of the ML metrics regarding the dopamine D1 receptor antagonist predictions using a dataset with reduced number of samples and without dimensionality reduction (GitHub file: https://github.com/articlesmli/NMR_ML_TTR_D1/blob/main/D1/4.1.ML_noPCA_3177.ipynb )

| 1.Algorithm | 2.Accuracy | 3.Precision | 4.Recall | 5.F1 | 6.ROC |
|---|---|---|---|---|---|
| SVM | 0.657 | 0.703 | 0.544 | 0.614 | 0.657 |
| GradientBoost | 0.650 | 0.692 | 0.541 | 0.607 | 0.650 |
| RandomForest | 0.601 | 0.822 | 0.259 | 0.394 | 0.601 |
| Decision | 0.571 | 0.588 | 0.471 | 0.523 | 0.571 |
| K-nearest | 0.512 | 0.520 | 0.309 | 0.387 | 0.512 |

Table ESM6. A five-fold cross-validation of ML model predicting the dopamine D1 receptor antagonists based on a reduce dataset samples and without dimensionality reduction (GitHub file: https://github.com/articlesmli/NMR_ML_TTR_D1/blob/main/D1/4.1.ML_noPCA_3177.ipynb )

| 1.Algorithm | 2.Mean CV Score | 3.Standard Deviation | 4.List of CV Scores |
|---|---|---|---|
| RandomForest | 0.8817 | 0.0495 | [0.847, 0.8083, 0.8856, 0.9242, 0.9435] |
| Decision | 0.7809 | 0.0334 | [0.7533, 0.7519, 0.7652, 0.7935, 0.8408] |
| SVM | 0.7753 | 0.0291 | [0.7504, 0.7385, 0.7816, 0.7845, 0.8214] |
| GradientBoost | 0.7045 | 0.0217 | [0.6805, 0.6805, 0.7058, 0.7207, 0.7351] |
| K-nearest | 0.6938 | 0.0287 | [0.6746, 0.6612, 0.6776, 0.7207, 0.7351] |

Table ESM7. Results of the ML metrics regarding the dopamine D1 receptor antagonist predictions using a dataset with reduced number of samples and PCA dimensionality reduction (GitHub file: https://github.com/articlesmli/NMR_ML_TTR_D1/blob/main/D1/4.2.ML_withPCA_3177.ipynb )

| 1.Algorithm | 2.Accuracy | 3.Precision | 4.Recall | 5.F1 | 6.ROC |
|---|---|---|---|---|---|
| SVM | 0.647 | 0.692 | 0.529 | 0.600 | 0.647 |
| GradientBoost | 0.600 | 0.705 | 0.344 | 0.462 | 0.600 |
| Decision | 0.537 | 0.559 | 0.350 | 0.430 | 0.537 |
| RandomForest | 0.534 | 0.677 | 0.129 | 0.217 | 0.534 |
| K-nearest | 0.519 | 0.532 | 0.321 | 0.400 | 0.519 |

Table ESM8. A five-fold cross-validation of the ML models predicting the dopamine D1 receptor antagonists based on data with reduced samples and PCA dimensionality reduction of the features (GitHub file: https://github.com/articlesmli/NMR_ML_TTR_D1/blob/main/D1/4.2.ML_withPCA_3177.ipynb )

| 1.Algorithm | 2.Mean CV Score | 3.Standard Deviation | 4.List of CV Scores |
|---|---|---|---|
| RandomForest | 0.8764 | 0.0565 | [0.8306, 0.7994, 0.8767, 0.9227, 0.9524] |
| Decision | 0.7854 | 0.0537 | [0.7801, 0.7266, 0.7296, 0.8276, 0.8631] |
| GradientBoost | 0.7747 | 0.0522 | [0.7281, 0.7073, 0.7831, 0.8024, 0.8527] |
| SVM | 0.7485 | 0.0362 | [0.7221, 0.6969, 0.7637, 0.7578, 0.8021] |
| K-nearest | 0.6917 | 0.0226 | [0.6672, 0.6657, 0.6969, 0.7043, 0.7247] |

Table ESM9 Metrics results of ML models based on a full dataset of 13CNMR spectroscopy data, predicting dopamine D1 receptor antagonists(GitHub file: https://github.com/articlesmli/NMR_ML_TTR_D1/blob/main/D1/6.1.ML_noPCA_fullSet.ipynb )

| 1.Algorithm | 2.Accuracy | 3.Precision | 4.Recall | 5.F1 | 6.ROC |
|---|---|---|---|---|---|
| SVM | 0.715 | 0.774 | 0.606 | 0.680 | 0.715 |
| XGBoost | 0.688 | 0.737 | 0.587 | 0.653 | 0.688 |
| RandomForest | 0.653 | 0.792 | 0.416 | 0.546 | 0.653 |
| GradientBoost | 0.650 | 0.674 | 0.582 | 0.625 | 0.650 |
| K-nearest | 0.612 | 0.671 | 0.441 | 0.532 | 0.612 |
| Decision | 0.578 | 0.596 | 0.483 | 0.534 | 0.578 |

Table ESM10. Five-fold cross-validation of ML based on a full dataset of 13CNMR spectroscopy data, predicting dopamine D1 receptor antagonists https://github.com/articlesmli/NMR_ML_TTR_D1/blob/main/D1/6.1.ML_noPCA_fullSet.ipynb )

| 1.Algorithm | 2.Mean CV Score | 3.Standard Deviation | 4.List of CV Scores |
|---|---|---|---|
| SVM | 0.7487 | 0.0030 | [0.7508, 0.7461, 0.7536, 0.7472, 0.7461] |
| XGBoost | 0.7207 | 0.0054 | [0.7252, 0.7144, 0.7236, 0.7265, 0.7141] |
| RandomForest | 0.7073 | 0.0021 | [0.7065, 0.7084, 0.7102, 0.7073, 0.704] |
| GradientBoost | 0.6885 | 0.0048 | [0.6925, 0.6872, 0.6861, 0.6954, 0.6817] |
| K-nearest | 0.6429 | 0.0059 | [0.6509, 0.6356, 0.6476, 0.6432, 0.637] |
| Decision | 0.6012 | 0.0065 | [0.6096, 0.6059, 0.5939, 0.6034, 0.5933] |

Table ESM11. Metric results of ML with a full dataset of 13CNMR data, dimensionality reduced by PCA for predicting dopamine D1 receptor antagonists (GitHub file: https://github.com/articlesmli/NMR_ML_TTR_D1/blob/main/D1/6.2..ML_withPCA_fullSet.ipynb)

| 1.Algorithm | 2.Accuracy | 3.Precision | 4.Recall | 5.F1 | 6.ROC |
|---|---|---|---|---|---|
| SVM | 0.642 | 0.696 | 0.506 | 0.586 | 0.642 |
| RandomForest | 0.619 | 0.704 | 0.411 | 0.519 | 0.619 |
| GradientBoost | 0.619 | 0.645 | 0.530 | 0.582 | 0.619 |
| XGBoost | 0.615 | 0.659 | 0.478 | 0.554 | 0.615 |
| K-nearest | 0.580 | 0.612 | 0.438 | 0.511 | 0.580 |
| Decision | 0.544 | 0.555 | 0.444 | 0.493 | 0.544 |

Table ESM12. Five-fold cross-validation of ML based on a full dataset with 13CNMR, reduced by PCA for predicting dopamine D1 receptor antagonists (GitHub file: https://github.com/articlesmli/NMR_ML_TTR_D1/blob/main/D1/6.2..ML_withPCA_fullSet.ipynb )

| 1.Algorithm | 2.Mean CV Score | 3.Standard Deviation | 4.List of CV Scores |
|---|---|---|---|
| SVM | 0.6830 | 0.0052 | [0.6881, 0.6808, 0.6782, 0.6903, 0.6778] |
| RandomForest | 0.6713 | 0.0072 | [0.6736, 0.6601, 0.6753, 0.6809, 0.6667] |
| GradientBoost | 0.6663 | 0.0074 | [0.6775, 0.659, 0.6654, 0.6718, 0.6581] |
| XGBoost | 0.6616 | 0.0040 | [0.6676, 0.6595, 0.6561, 0.6645, 0.6604] |
| K-nearest | 0.6205 | 0.0043 | [0.6239, 0.6147, 0.6158, 0.6239, 0.6242] |
| Decision | 0.5780 | 0.0094 | [0.5878, 0.5752, 0.5856, 0.5801, 0.5615] |

Table ESM13. Metric results of ML models, predicting TTR transcription antagonists based on a dataset with added molecular features (GitHub file: https://github.com/articlesmli/NMR_ML_TTR_D1/blob/main/TTR/5.1.ML_noPCA_withPubChem_data.ipynb )

| 1.Algorithm | 2.Accuracy | 3.Precision | 4.Recall | 5.F1 | 6.ROC |
|---|---|---|---|---|---|
| GradientBoost | 0.671 | 0.740 | 0.526 | 0.615 | 0.671 |
| RandomForest | 0.656 | 0.794 | 0.421 | 0.550 | 0.656 |
| SVM | 0.604 | 0.671 | 0.409 | 0.508 | 0.604 |
| Decision | 0.596 | 0.638 | 0.441 | 0.522 | 0.596 |
| K-nearest | 0.547 | 0.562 | 0.429 | 0.487 | 0.547 |

Table ESM14. A five-fold cross-validation for predicting TTR transcription antagonists based on a dataset with added molecular features (GitHub file: https://github.com/articlesmli/NMR_ML_TTR_D1/blob/main/TTR/5.1.ML_noPCA_withPubChem_data.ipynb )

| 1.Algorithm | 2.Mean CV Score | 3.Standard Deviation | 4.List of CV Scores |
|---|---|---|---|
| RandomForest | 0.8899 | 0.0369 | [0.8556, 0.8507, 0.8756, 0.9331, 0.9347] |
| Decision | 0.8032 | 0.0429 | [0.75, 0.7792, 0.7792, 0.86, 0.8476] |
| SVM | 0.7245 | 0.0272 | [0.6879, 0.7045, 0.7201, 0.7496, 0.7605] |
| GradientBoost | 0.6931 | 0.0186 | [0.6599, 0.6874, 0.6983, 0.7123, 0.7076] |
| K-nearest | 0.6645 | 0.0499 | [0.6335, 0.6112, 0.6345, 0.7465, 0.6967] |

Table ESM15. ML metric results of ML models predicting TTR transcription activators based on a dataset with 13C NMR and molecular feature and dimensionality reduced by PCA (GitHub file: https://github.com/articlesmli/NMR_ML_TTR_D1/blob/main/TTR/5.2..ML_withPCA_withPubChem_data.ipynb )

| 1.Algorithm | 2.Accuracy | 3.Precision | 4.Recall | 5.F1 | 6.ROC |
|---|---|---|---|---|---|
| SVM | 0.650 | 0.707 | 0.512 | 0.594 | 0.650 |
| GradientBoost | 0.604 | 0.625 | 0.521 | 0.568 | 0.604 |
| K-nearest | 0.534 | 0.536 | 0.503 | 0.519 | 0.534 |
| Decision | 0.519 | 0.528 | 0.359 | 0.427 | 0.519 |
| RandomForest | 0.519 | 0.528 | 0.359 | 0.427 | 0.519 |

Table ESM16. A five-fold cross-validation for predicting TTR transcription activators based on a dataset with 13C NMR and molecular feature and dimensionality reduced by PCA (GitHub file: https://github.com/articlesmli/NMR_ML_TTR_D1/blob/main/TTR/5.2..ML_withPCA_withPubChem_data.ipynb )

| 1.Algorithm | 2.Mean CV Score | 3.Standard Deviation | 4.List of CV Scores |
|---|---|---|---|
| GradientBoost | 0.7172 | 0.0127 | [0.7061, 0.7023, 0.7138, 0.7352, 0.7286] |
| RandomForest | 0.7146 | 0.0099 | [0.7126, 0.7072, 0.7023, 0.7303, 0.7204] |
| SVM | 0.6952 | 0.0093 | [0.6847, 0.6891, 0.6908, 0.7007, 0.7105] |
| K-nearest | 0.6449 | 0.0203 | [0.6338, 0.6382, 0.6266, 0.6414, 0.6842] |
| Decision | 0.6090 | 0.0092 | [0.5944, 0.6102, 0.6168, 0.6036, 0.6201] |

Table ESM17. ML metric results based on a reduced dataset of 13CNMR and molecular features for predicting dopamine D1 receptor antagonists (GitHub file: https://github.com/articlesmli/NMR_ML_TTR_D1/blob/main/D1/5.1.ML_noPCA_3177_withPubChem_data.ipynb )

| 1.Algorithm | 2.Accuracy | 3.Precision | 4.Recall | 5.F1 | 6.ROC |
|---|---|---|---|---|---|
| GradientBoost | 0.751 | 0.835 | 0.626 | 0.716 | 0.751 |
| RandomForest | 0.746 | 0.871 | 0.576 | 0.694 | 0.746 |
| SVM | 0.690 | 0.774 | 0.535 | 0.633 | 0.690 |
| Decision | 0.659 | 0.694 | 0.568 | 0.625 | 0.659 |
| K-nearest | 0.594 | 0.654 | 0.400 | 0.496 | 0.594 |

Table ESM18. Five-fold cross-validation of ML based on a reduced dataset with 13CNMR and molecular features for predicting dopamine D1 receptor antagonists (GitHub file: https://github.com/articlesmli/NMR_ML_TTR_D1/blob/main/D1/5.1.ML_noPCA_3177_withPubChem_data.ipynb )

| 1.Algorithm | 2.Mean CV Score | 3.Standard Deviation | 4.List of CV Scores |
|---|---|---|---|
| RandomForest | 0.9138 | 0.0287 | [0.9094, 0.8692, 0.9004, 0.945, 0.9449] |
| Decision | 0.8404 | 0.0303 | [0.8009, 0.8083, 0.8499, 0.8707, 0.872] |
| SVM | 0.8166 | 0.0190 | [0.7964, 0.792, 0.8262, 0.8276, 0.8408] |
| GradientBoost | 0.7878 | 0.0128 | [0.7875, 0.7637, 0.792, 0.8009, 0.7946] |
| K-nearest | 0.7331 | 0.0507 | [0.6835, 0.6627, 0.7533, 0.7771, 0.7887] |

Table ESM19.ML metric results based on ML based on a reduced dataset with 13CNMR data and molecular features, dimensionality reduced by PCA for predicting dopamine D1 receptor antagonists (GitHub file: https://github.com/articlesmli/NMR_ML_TTR_D1/blob/main/D1/7.2..ML_withPCA_fullSet_withPubChem_data.ipynb )

| 1.Algorithm | 2.Accuracy | 3.Precision | 4.Recall | 5.F1 | 6.ROC |
|---|---|---|---|---|---|
| GradientBoost | 0.704 | 0.833 | 0.512 | 0.634 | 0.704 |
| SVM | 0.687 | 0.759 | 0.547 | 0.636 | 0.687 |
| RandomForest | 0.672 | 0.868 | 0.406 | 0.553 | 0.672 |
| Decision | 0.585 | 0.626 | 0.424 | 0.505 | 0.585 |
| K-nearest | 0.576 | 0.621 | 0.391 | 0.480 | 0.576 |

Table ESM20. Five-fold cross-validation of ML based on a reduced dataset with 13CNMR data and molecular features, dimensionality reduced by PCA for predicting dopamine D1 receptor antagonists (GitHub file: https://github.com/articlesmli/NMR_ML_TTR_D1/blob/main/D1/7.2..ML_withPCA_fullSet_withPubChem_data.ipynb )

| 1.Algorithm | 2.Mean CV Score | 3.Standard Deviation | 4.List of CV Scores |
|---|---|---|---|
| RandomForest | 0.9031 | 0.0378 | [0.8618, 0.8692, 0.8886, 0.9421, 0.9539] |
| GradientBoost | 0.8347 | 0.0264 | [0.7994, 0.8187, 0.8247, 0.8648, 0.8661] |
| Decision | 0.8220 | 0.0352 | [0.7771, 0.789, 0.8217, 0.8559, 0.8661] |
| SVM | 0.8121 | 0.0240 | [0.7845, 0.7979, 0.7994, 0.8276, 0.8512] |
| K-nearest | 0.7307 | 0.0522 | [0.6835, 0.6761, 0.7073, 0.7979, 0.7887] |

Table ESM21. ML metric results based on a full dataset of 13CNMR and molecular features for predicting dopamine D1 receptor antagonists (GitHub file: https://github.com/articlesmli/NMR_ML_TTR_D1/blob/main/D1/7.1.ML_noPCA_fullSet_withPubChem_data.ipynb )

| 1.Algorithm | 2.Accuracy | 3.Precision | 4.Recall | 5.F1 | 6.ROC |
|---|---|---|---|---|---|
| GradientBoost | 0.758 | 0.842 | 0.636 | 0.724 | 0.758 |
| RandomForest | 0.756 | 0.872 | 0.600 | 0.711 | 0.756 |
| SVM | 0.747 | 0.822 | 0.631 | 0.714 | 0.747 |
| Decision | 0.687 | 0.715 | 0.621 | 0.665 | 0.687 |
| K-nearest | 0.615 | 0.657 | 0.483 | 0.556 | 0.615 |

Table ESM22. Fi-fold cross-validation of ML based on a full dataset with 13CNMR and molecular features for predicting dopamine D1 receptor antagonists (GitHub file: https://github.com/articlesmli/NMR_ML_TTR_D1/blob/main/D1/7.1.ML_noPCA_fullSet_withPubChem_data.ipynb )

| 1.Algorithm | 2.Mean CV Score | 3.Standard Deviation | 4.List of CV Scores |
|---|---|---|---|
| RandomForest | 0.8878 | 0.0469 | [0.8475, 0.8473, 0.8547, 0.9362, 0.9532] |
| Decision | 0.8085 | 0.0400 | [0.7715, 0.7781, 0.7783, 0.8521, 0.8623] |
| SVM | 0.8007 | 0.0210 | [0.7785, 0.7799, 0.7946, 0.8218, 0.8288] |
| GradientBoost | 0.7607 | 0.0039 | [0.7599, 0.7548, 0.7636, 0.7589, 0.7662] |
| K-nearest | 0.7104 | 0.0391 | [0.6742, 0.6812, 0.6804, 0.7599, 0.7564] |

Table ESM23. Metrics results of ML with a full dataset of 13CNMR and molecular features and dimensionality reduced by PCA for predicting dopamine D1 receptor antagonists (GitHub file: https://github.com/articlesmli/NMR_ML_TTR_D1/blob/main/D1/7.2..ML_withPCA_fullSet_withPubChem_data.ipynb )

| 1.Algorithm | 2.Accuracy | 3.Precision | 4.Recall | 5.F1 | 6.ROC |
|---|---|---|---|---|---|
| SVM | 0.743 | 0.820 | 0.623 | 0.708 | 0.743 |
| GradientBoost | 0.739 | 0.809 | 0.624 | 0.705 | 0.739 |
| RandomForest | 0.719 | 0.847 | 0.534 | 0.655 | 0.719 |
| Decision | 0.657 | 0.689 | 0.572 | 0.625 | 0.657 |
| K-nearest | 0.620 | 0.662 | 0.491 | 0.564 | 0.620 |

Table ESM24. Five-fold cross-validation of ML based on a full dataset with 13CNMR and molecular feature and dimensionality reduced by PCA for predicting dopamine D1 receptor antagonists (GitHub file: https://github.com/articlesmli/NMR_ML_TTR_D1/blob/main/D1/7.2..ML_withPCA_fullSet_withPubChem_data.ipynb )

| 1.Algorithm | 2.Mean CV Score | 3.Standard Deviation | 4.List of CV Scores |
|---|---|---|---|
| RandomForest | 0.8701 | 0.0509 | [0.8277, 0.8249, 0.8337, 0.925, 0.9393] |
| SVM | 0.7931 | 0.0173 | [0.7748, 0.7758, 0.7904, 0.804, 0.8204] |
| Decision | 0.7840 | 0.0447 | [0.746, 0.7482, 0.7489, 0.8306, 0.8463] |
| GradientBoost | 0.7512 | 0.0052 | [0.7474, 0.7439, 0.7538, 0.7521, 0.7589] |
| K-nearest | 0.7056 | 0.0382 | [0.6714, 0.6734, 0.679, 0.7472, 0.7571] |

Table ESM25. CID_SID ML model`s metrics predicting TTR transcription activation capability of compounds designed initially for another purpose (GitHub file: https://github.com/articlesmli/NMR_ML_TTR_D1/blob/main/TTR_CID_SID_ML_model/TTR_CID_SID_ML_model.ipynb )

| 1.Algorithm | 2.Accuracy | 3.Precision | 4.Recall | 5.F1 | 6.ROC |
|---|---|---|---|---|---|
| GradientBoost | 0.815 | 0.946 | 0.668 | 0.783 | 0.815 |
| K-nearest | 0.791 | 0.844 | 0.715 | 0.774 | 0.791 |
| RandomForest | 0.785 | 0.844 | 0.700 | 0.765 | 0.785 |
| SVM | 0.774 | 0.970 | 0.565 | 0.714 | 0.774 |
| Decision | 0.740 | 0.774 | 0.676 | 0.722 | 0.740 |

Table ESM26. Five-fold cross-validation of CID_SID ML models predicting TTR transcription activation capability of compounds that have been initially designed for another purpose (GitHub file: https://github.com/articlesmli/NMR_ML_TTR_D1/blob/main/TTR_CID_SID_ML_model/TTR_CID_SID_ML_model.ipynb )

| 1.Algorithm | 2.Mean CV Score | 3.Standard Deviation | 4.List of CV Scores |
|---|---|---|---|
| GradientBoost | 0.8458 | 0.0085 | [0.8472, 0.852, 0.8297, 0.847, 0.8533] |
| RandomForest | 0.8301 | 0.0055 | [0.8205, 0.8362, 0.8312, 0.8281, 0.8344] |
| K-nearest | 0.8175 | 0.0067 | [0.8205, 0.8157, 0.8076, 0.8155, 0.8281] |
| SVM | 0.8108 | 0.0139 | [0.8079, 0.8205, 0.8155, 0.7855, 0.8249] |
| Decision | 0.7799 | 0.0078 | [0.7811, 0.7874, 0.7681, 0.7744, 0.7886] |

# Figures

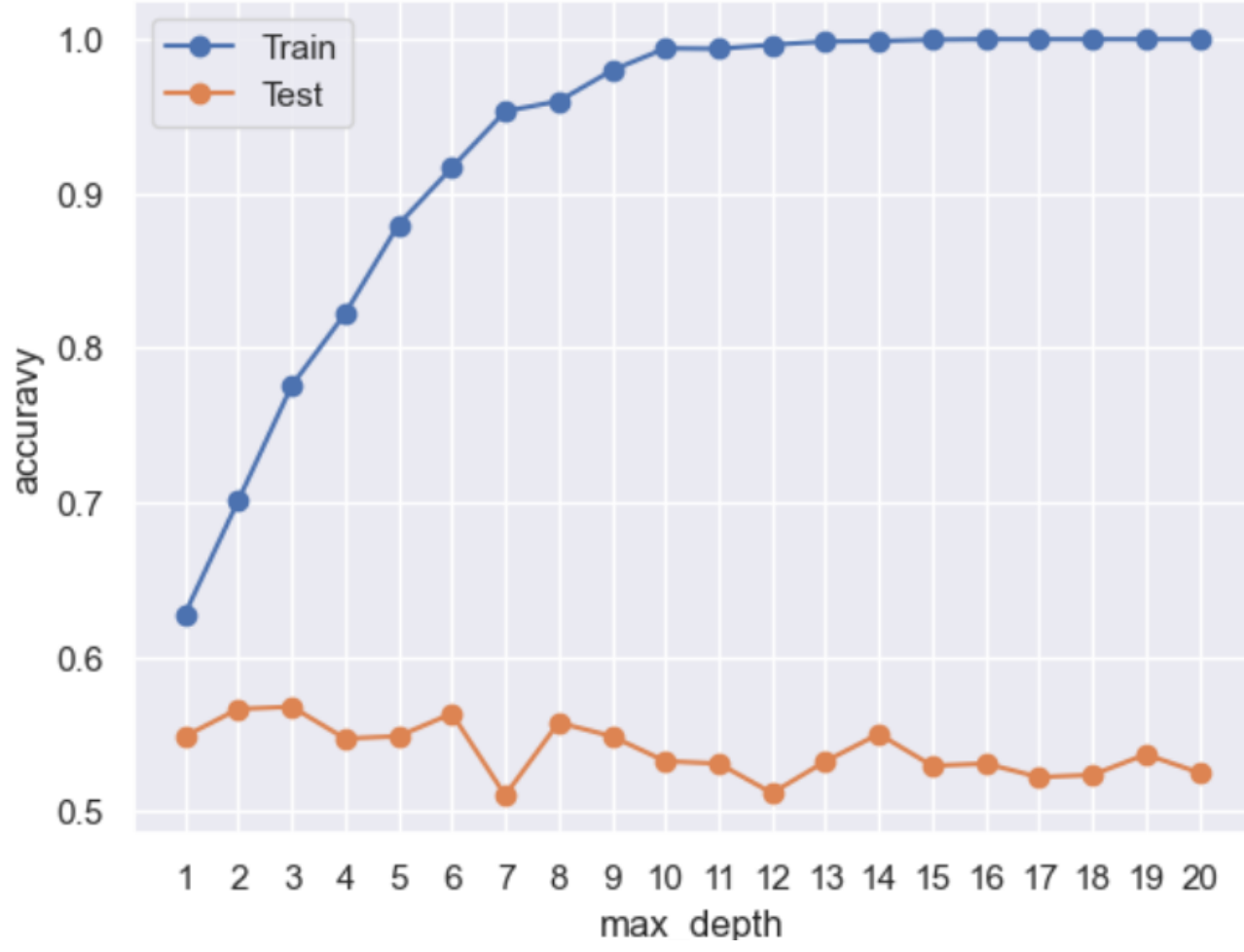

Figure ESM1. Tracing the deviation between train and test accuracies for the TTR case without dimensionality reduction

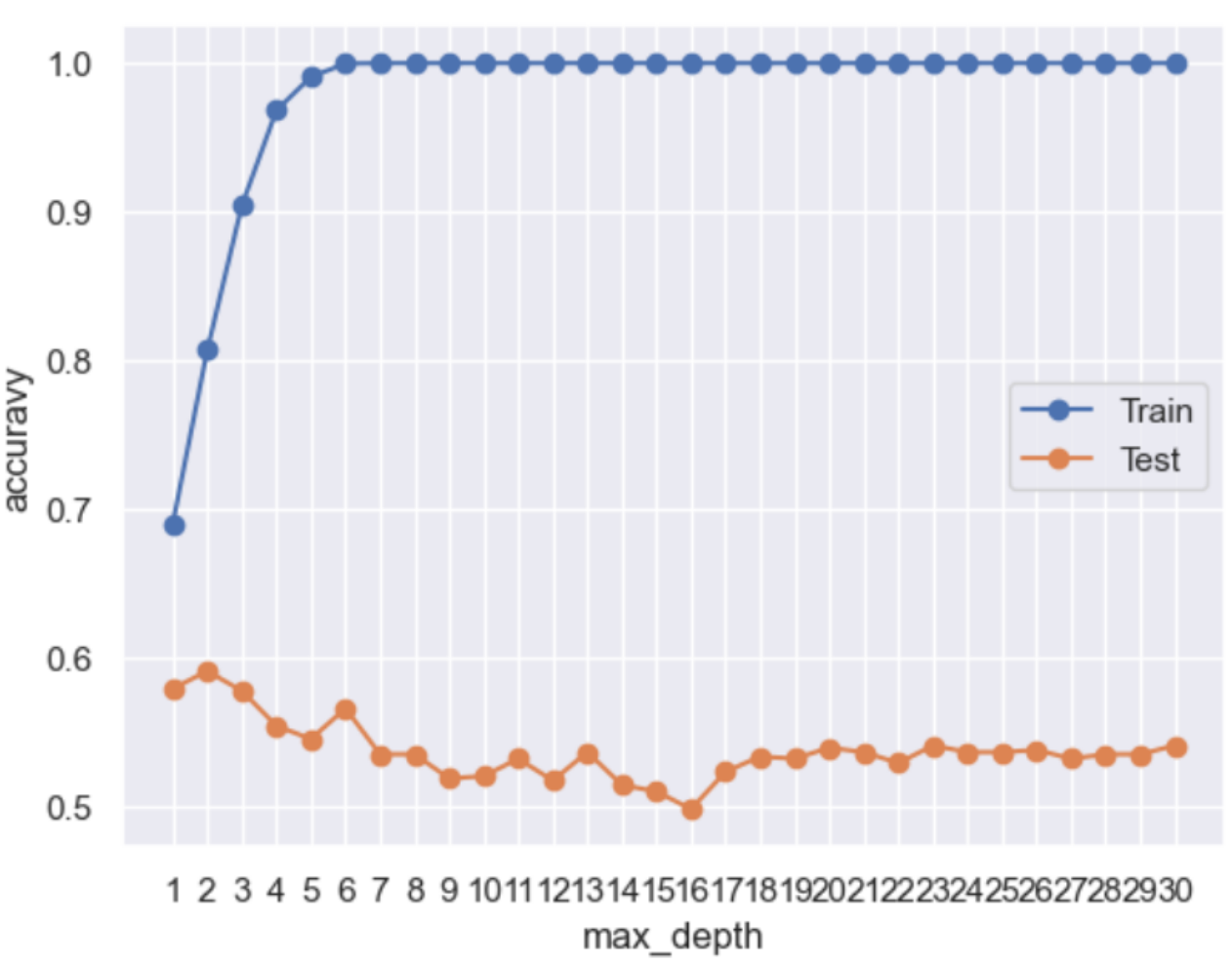

Figure ESM2. Tracing the deviation between train and test accuracies for the TTR case with dimensionality reduction performed by PCA

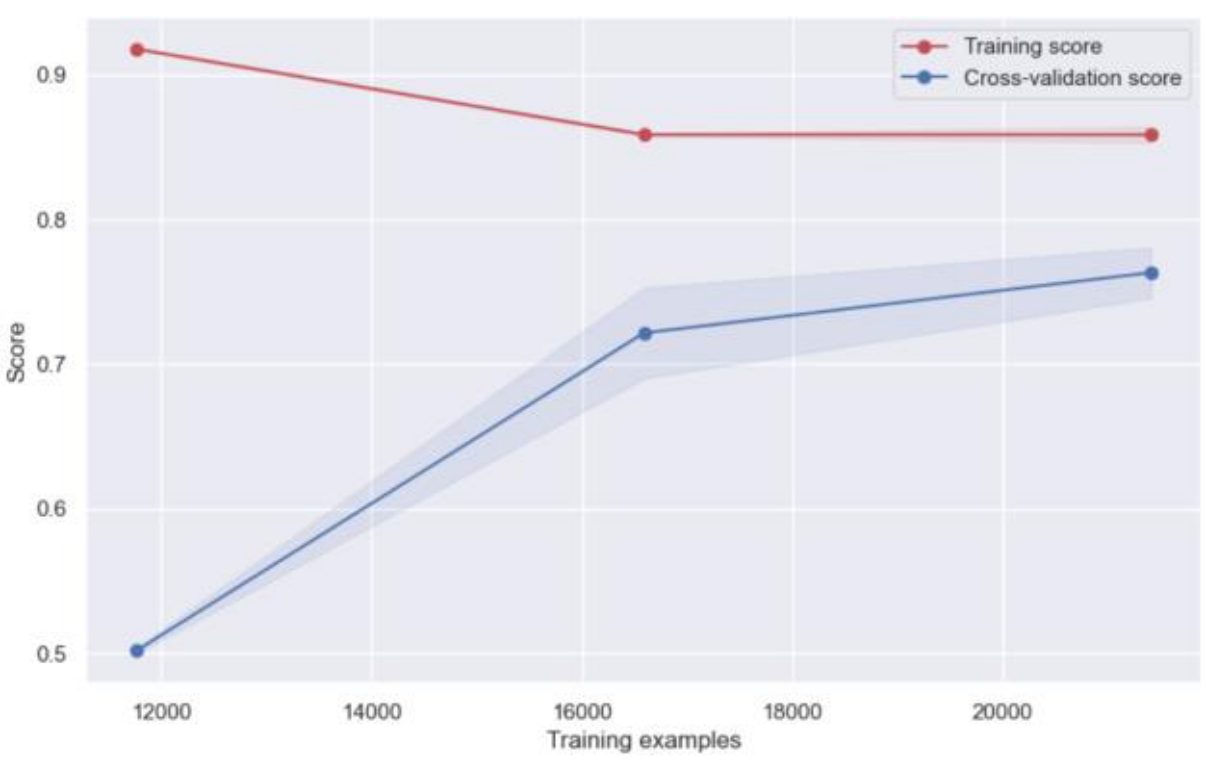

Figure ESM3. Learning curve dopamine D1 receptor case without dimensionality reduction

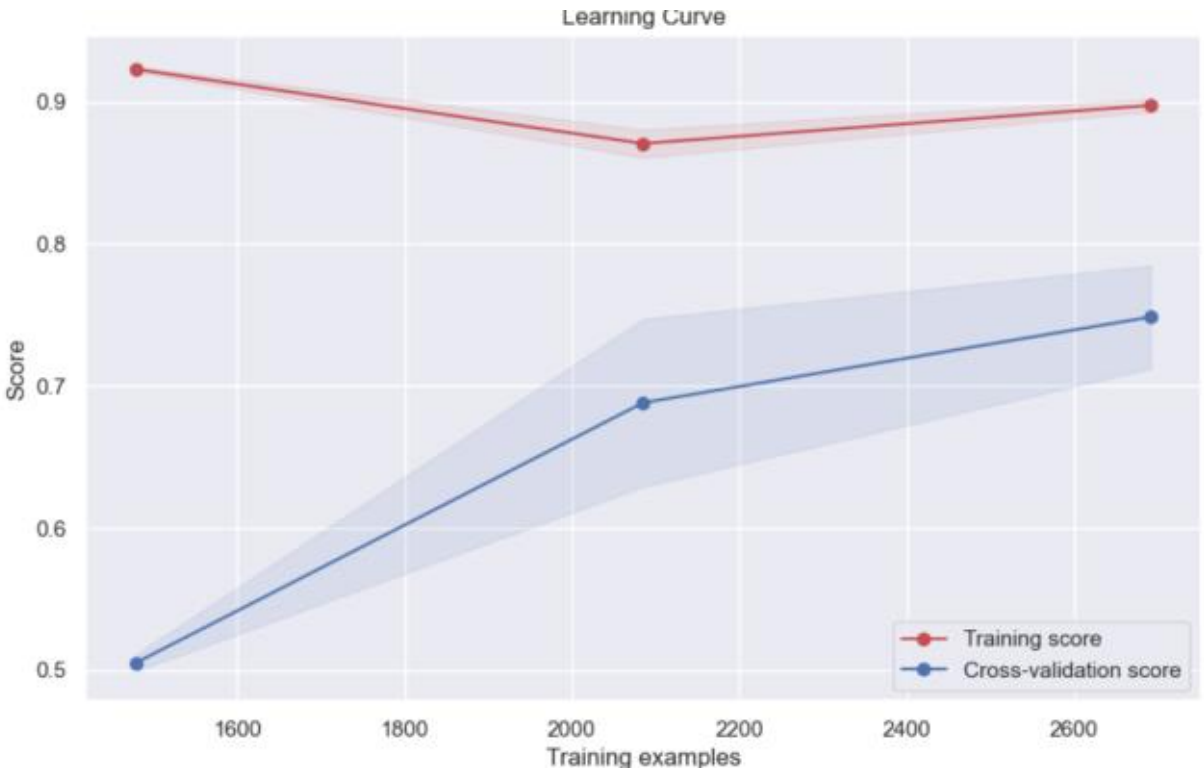

Figure ESM4. Learning curve of ML model predicting dopamine D1 receptor antagonists with dimensionality reduction performed by PCA

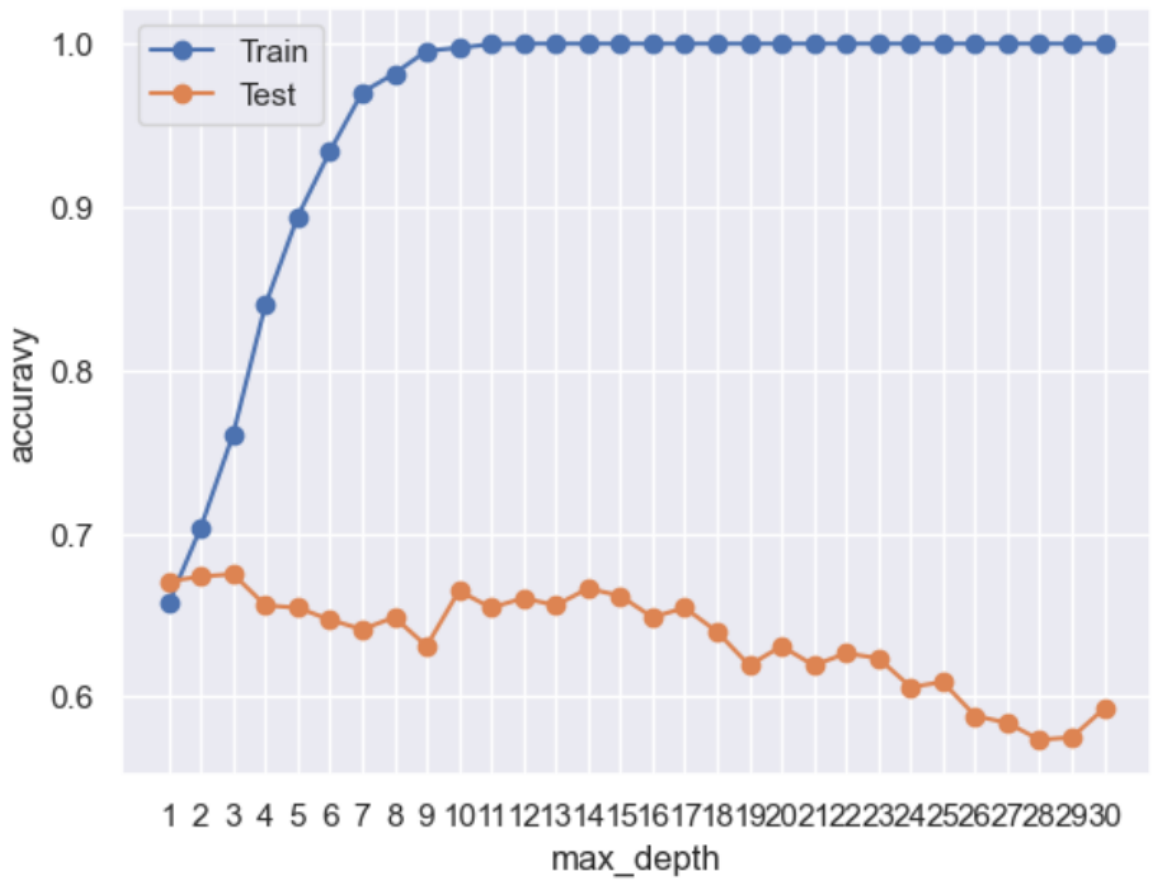

Figure ESM5. Tracing the deviation between train and test accuracies for the TTR case with molecular features added without dimensionality reduction

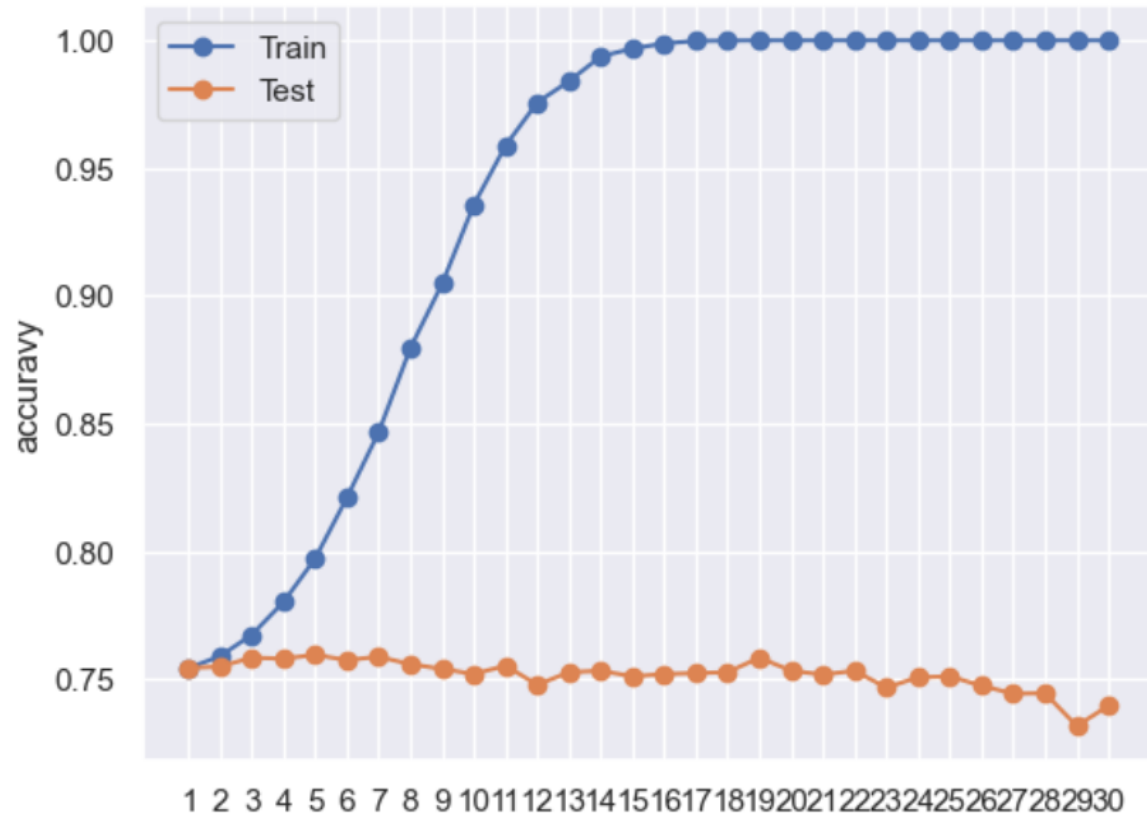

Figure ESM6. Tracing the deviation between train and test accuracies for the dopamine D1 receptor antagonist and molecular features case without dimensionality reduction

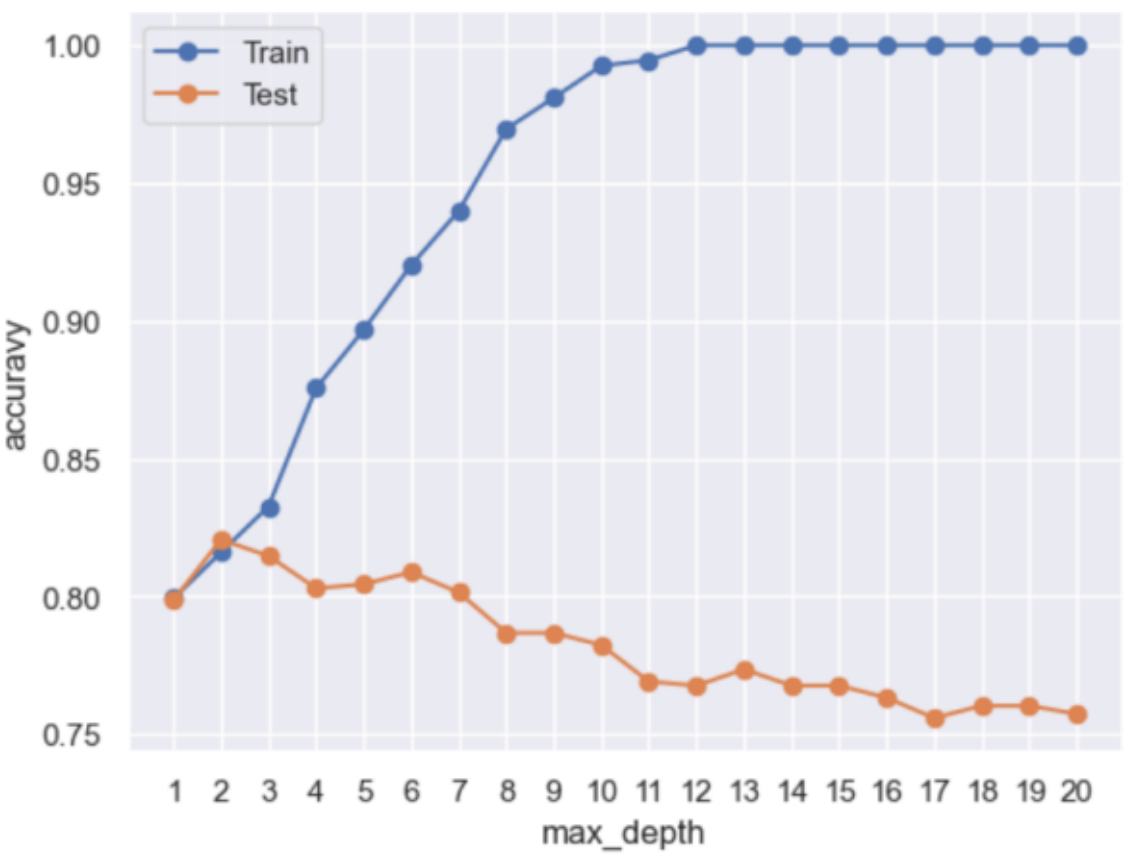

Figure ESM7. Tracing the deviation between train and test accuracies of the CID_SID ML model based on a dataset focused on predicting TTR transcription activators.